# Clear, Concise and Effective UI: Opinion and Suggestions

Rishabh Jain
Rupanta Rwiteej Dutta
Rajat Tandon

2Abstract

The most important aspect of any Software is the operability for the intended audience. This factor of operability is encompassed in the user interface, which serves as the only window to the features of the system. It is thus essential that the User Interface provided is robust, concise and lucid. Presently there are no properly defined rules or guidelines for user interface design enabling a 'perfect' design, since such a system cannot be perceived. This article aims at providing suggestions in the design of the User Interface, which would make it easier for the user to navigate through the system features and also the developers to guide the users towards better utilization of the features.

Rishabh Jain

Rupanta Rwiteej Dutta

Rajat Tandon

7 September 2014

Clear, Concise and Effective UI: Opinion and Suggestions

The evolution of the User Interface began with the advent of simple applications, which used Command Line Interface. This interface required the user to know the commands that would serve the intended purpose, which required the user to have prior knowledge of the working of the system. Since at that time the use of computers was not very widespread, primarily restricted to amateur programmers and research laboratories, having a learnable User Interface was not a motivation. With the increased use of computation devices have led the masses towards a more extended interaction with software, driving applications to have an easily foreseeable interface. This requirement is subject to audience notions and preferences, leading to ambiguity about the design of the User Interface.

"Good design is obvious. Great design is transparent."[1]
It is reckonable that there cannot be a perfect UI per se, but there is a skeletal idea about an ideal User Interface, which has the certain characteristics:
- Responsiveness
- Familiarity

- Clarity
- Conciseness

To maintain these characteristics, we suggest that User Interface designers follow the following guidelines.

**Responsive Web Design** has erupted as a result of high definition displays. Since these displays support high pixel count with resolutions up to 1920x1080, it enables the developers to display a large amount of content at a time. But the adaptation to these displays has not been speedy, with a lot of users still using primitive resolution displays. It is recommended that the user interface should be rendered adaptively across a range of resolutions, without being disintegrated. A best-viewed resolution may be in place, however, the position of various screen elements relative to one another, should not be affected by a change in screen size. The design of the UI should fluidly change and respond to the screen size.

**Creating visually appealing content** is essential to swiftly and strategically attract the attention of the User. Hierarchical and logically binding order of content is an integral part of a good idea. The user should easily be able to distinguish elements of higher significance or value from the less important ones. Hierarchical ordering also generates friction, generating room for the user to individually analyze the content rather than skimming through it. Combining content ordering with other styles such as font size, font color, element size, padding, margin, spacing, alignment, etc. creates a more visually appealing as well as logically sequenced experience to the user.

**Having a single columned layout** instead of having multiple columns is preferred since it gives more control in guiding a reader in a more foreseeable way from the beginning to the end. Contrast to the above mentioned method, is to having multiple column, which has the additional risk of being distracting to the core purpose of the page.

**Protuberant and differentiable action calls** help the user to identify the appropriate response and is an important characteristic of a good UI. The action calls can be buttons, links or any other element, which can trigger a particular action. In order to make them distinguishable, the developer may use a darker color shade over a light colored UI or vice versa. He might even adjust the element's depth by giving it drop-shadows and gradients, to make it look closer while the rest of the content appears to be far away.

**Categorizing similar features, functions and actions** and grouping them together, is a good practice. Disordering of similar content increases the time required to complete a certain task. The user finds it difficult accomplish his objective. Associated items should be placed near to each other in order to respect a degree of logic and lower the overall required cognitive function of the human brain.

**Implementing pagination while handling huge data** helps users to organize thoughts and makes it easier to swiftly reach the logical destination in the content. For instance, implementing a scroll helps to display content of size more than the actual size of the screen. However, having a very long scroll is not a good practice. In case of having a long scroll, a user may want to return to the top content by the time he reaches the bottom of the page to reiterate. Another instance can be the case, where a user at one end of the page, wishes to move to the other end; he or she has to scroll a long way, before arriving at the desired content.

A simple solution to this problem is to implement pagination, or categorizing and dividing the entire content into different pages and thereafter applying scroll individually.

**Form validation should be made inline** rather than calling the validation function, at the end of the process, when the Save/Submit button is clicked. In case the validation method being called at the end and error message is displayed subsequently, the user is required to go back and correct the invalid entry he had previously committed. This increases both the feature usage turnaround time as well as user fatigue. Whereas, the inline validation method shows an error message just when the user commits the entry and requires him to fix it immediately reducing both the incorrect entry lookup time as well as correction time.

**Displaying item states** helps the user understand whether his past actions have been successfully carried out or whether any action needs to be taken by providing him with appropriate feedback. The feedback presents the current state of the concerned item to the user. Examples of such feedback can be Bills paid/unpaid, Emails read/unread, any Option or Feature selected/unselected, etc.

**Color themes** play an important role in the look and feel of any good UI and should be selected appropriately according to the targeted audience. It shouldn't be too flashy neither should be dull. The theme should also adhere to the central idea of the web page/application. For instance, the color red may be associated with energy, positivity, war, action, danger, strength, courage and all things that are intense and passionate. Whereas, the color blue is associated with the sea, the sky, trust, honesty, loyalty, sincere, peace, tranquility and intelligence. The driving motivation here should be associating the color theme with the overall user emotion.

**Transition or Animation** to the interface elements, as a feedback to user interactions might be useful occasionally as part of the interface. For instance, it is a common practice to display promotional information in motion on the screen. However, while implementing it, the timing for transition needs to be uniform and appropriate. The delay should be long enough to enable the

8user to perceive the information as well as quick enough to successfully engage the user. Also the content should comprise of more graphical data, which is easy to perceive, than just text.

A built in intentional delay in the form of an animation or transition makes it easier to comprehend the action appropriated. It respects cognition and gives the user required time to understand a change in size or position.

**Refactoring UI to avoid duplication** is essential since over prolonged periods of UI development, it is possible to unintentionally create fragmented UI components, which all perform the same function. These fragmented UI components can be multiple sections, elements (buttons and links) and features. Having a replicate functionality, labeled in different ways, puts a strain on the users, thereby towering the learning curve. Refactoring the UI occasionally, by combining similar functions together may prove to be fruitful.

Lastly, **Object Oriented Approach to UI Development** should be implemented as a measure for the developers. Classes should be used in order to handle common styles across various screens, instead of having inline styles for each of them. This practice makes code management much easier. Just by changing a value at one particular location, the change in element property can be reflected across multiple screens.

While following the opinions and suggestions in this article, it must be reminded that the quality of the UI is subject to individual preferences of the user and cannot be molded perfectly. The intention of the developer should be to most closely replicate the desired interface for the majority of the users and enable the most accurate and efficient usage of the features.

Cited References